\documentclass[a4paper,twocolumn]{quantumarticle}
\pdfoutput=1
\usepackage{amsthm}
\usepackage{amsfonts}
\usepackage{pgf}
\usepackage{tikz}
\usetikzlibrary{matrix,fit,backgrounds,positioning}
\usepackage{relsize}
\usepackage{color}
\usepackage[utf8]{inputenc}
\usepackage{capt-of}
\usepackage{mathtools}
\usepackage[numbers,sort&compress]{natbib}
\usepackage{float}
\usepackage[section]{placeins}
\usepackage{listings}
\usepackage[T1]{fontenc} 
\usetikzlibrary{decorations.pathreplacing}

\usepackage{graphicx,amsfonts,bm,amsmath,amssymb,verbatim,ifthen,braket,xcolor,array,comment,enumerate,multirow,soul,lineno,bbm,array,lmodern}
\usepackage[normalem]{ulem}

\usepackage[version=4]{mhchem}
\usepackage[english]{babel}
\makeatletter
\adddialect\l@en\l@english
\makeatother
\usepackage[rightcaption]{sidecap}

\usepackage{booktabs}
\usepackage{longtable}
\usepackage{tabularx}
\usepackage{fvextra}
\usepackage[colorlinks=true,citecolor=blue,linkcolor=blue,urlcolor=blue]{hyperref}
\usepackage[all]{hypcap}

\DefineVerbatimEnvironment{jsonverbatim}{Verbatim}{
  fontsize=\scriptsize,
  breaklines=true,
  breakanywhere=true,
  frame=single,
  breaksymbolleft={},   
  breaksymbolright={}  
}
\DeclareUnicodeCharacter{2009}{\,}

\begin{document}

\title{Scalable Measurement-Based Quantum Simulation Patterns for Benchmarking}

\author{V. W. Scarola}
\email[Email address:]{scarola@vt.edu}
\affiliation{Department of Physics, Virginia Tech, Blacksburg, Virginia 24061, USA}

\begin{abstract}
Measurement-based quantum computing uses measurement patterns on predefined quantum resource states to execute quantum logic. Quantum simulation offers an important use case on near-term devices. However, pattern optimization depends on the multivariable interplay between hardware and software constraints and is therefore use-dependent and highly non-trivial. Optimization of large-scale patterns under realistic assumptions remains a barrier. We announce the release of the quantum measurement pattern library QPatLib, a dataset that, in v1.0, presents patterns for use in measurement-based quantum simulation. We present the workflow for generating patterns that execute Pauli-string unitaries needed for many quantum algorithms. We provide benchmark patterns for measurement-based quantum unitary evolution. The measurement patterns are defined with different conventions for commuting Pauli-string subsets to allow scaling of pattern size and complexity. The purpose of the library is to (i) serve as a standardized testbed for pattern-optimization protocols for measurement-based quantum simulation routines, (ii) offer a suite of patterns for direct use on hardware, (iii) provide data to empirically justify pattern design principles, and (iv) provide a flexible resource for future storage and use of measurement-based patterns beyond quantum simulation.
\end{abstract}

\emph{The QPatLib v1.0 datasets are available at \href{https://doi.org/10.5281/zenodo.20115266}{doi.org/10.5281/zenodo.20115266}}

\maketitle

\section{Introduction}

The measurement-based quantum computing (MBQC) approach executes quantum logic operations with adaptive local projective measurements on predefined quantum resource states \cite{RAUSSENDORF2001,RAUSSENDORF2003,WEI2021}. Certain families of graph states (notably 2D cluster states and related constructions) built from nodes (physical qubits) and edges (entangling gates) offer universal quantum resource states \cite{HEIN2006}. All quantum algorithms can be performed by executing measurement \emph{patterns} on suitable universal graph states.

Quantum simulation algorithms, such as variational quantum eigensolvers \cite{ROMERO2019,CEREZO2021a,BHARTI2022} and algorithms in the quantum phase estimation family \cite{Kitaev1995,Abrams1999,Ortiz2001,Somma2002,ASPURU-GUZIK2005a,Poulin2018}, offer excellent candidates for measurement-based quantum simulation (MBQS). For example, recent work \cite{MOTTA2020,SHEN2023a,YU2025a,BHARTI2021,HAUG2022,LEE2024,ZHAO2025,LEE2025a} builds hybrid classical/quantum algorithms in the circuit approach that aim to lower overhead for quantum phase estimation--like algorithms, making them strong near-term use cases for MBQS. As a concrete example, a noise-resilient hybrid quantum gap-estimation algorithm \cite{LEE2022,LEE2024,LEE2025a} uses quantum time evolution on static many-body Hamiltonians to extract targeted energy gaps with classical feedback and an offline (classical) Fourier transform. The compact nature of such hybrid algorithms implies that they can, in principle, be scaled up on quantum devices as hardware improves.

Recent work shows advantages in implementing quantum simulation algorithms either entirely or partly in the MBQS framework \cite{FERGUSON2021a,LEE2022,MARQVERSEN2023,CHAN2024,KALDENBACH2025,KALDENBACH2025a}. A central primitive in these approaches is the exponential of Pauli strings, as required for Hamiltonian time evolution when many-body models are expressed as sums of Pauli strings \cite{Ortiz2001,Somma2002,Whitfield2011}. Classically intractable fermionic systems \cite{TROYER2005}, including Hubbard and quantum chemistry Hamiltonians \cite{MCARDLE2020a,BHARTI2022,DALEY2022}, therefore provide natural benchmarks for MBQS.

Optimizing MBQS patterns that implement Pauli-string exponentials is challenging: compactifying the underlying graph state is itself a non-trivial many-body problem \cite{VANDENNEST2004c,HEIN2006,BROADBENT2009a,DUNCAN2010,BROWNE2011b,DAHLBERG2018a,CLAUDET2025b,KALDENBACH2025,KALDENBACH2025a,SHARMA2026}. Moreover, meaningful optimization must account for the measurement pattern as a whole \cite{BROADBENT2009a}, since practical performance depends jointly on graph structure and measurement scheduling. Representative goals include minimizing edge ``lifetime'', width, and degree while supporting repeatable pattern structure \cite{KALDENBACH2025a}.

Crucially, these constraints are not universal: they depend on hardware connectivity and on the fault-tolerant substrate. For example, trapped-ion platforms can offer highly flexible (including all-to-all) entangling operations, whereas superconducting architectures often face planar connectivity constraints such as nearest-neighbor coupling \cite{BRUZEWICZ2019,KJAERGAARD2020} and photonic platforms can favor chain connectivity \cite{LINDNER2009, SCHWARTZ2016,ISTRATI2020}. Likewise, measurement-based fault tolerance can impose different resource-state requirements, e.g., three-dimensional versus two-dimensional cluster-state constructions \cite{Raussendorf2006,Raussendorf2007b}. This diversity of constraints means there is no single ``best'' pattern independent of use case, motivating a standard dataset for controlled comparisons.

\begin{figure}[t]
\centering 
\includegraphics[width=0.45\textwidth]{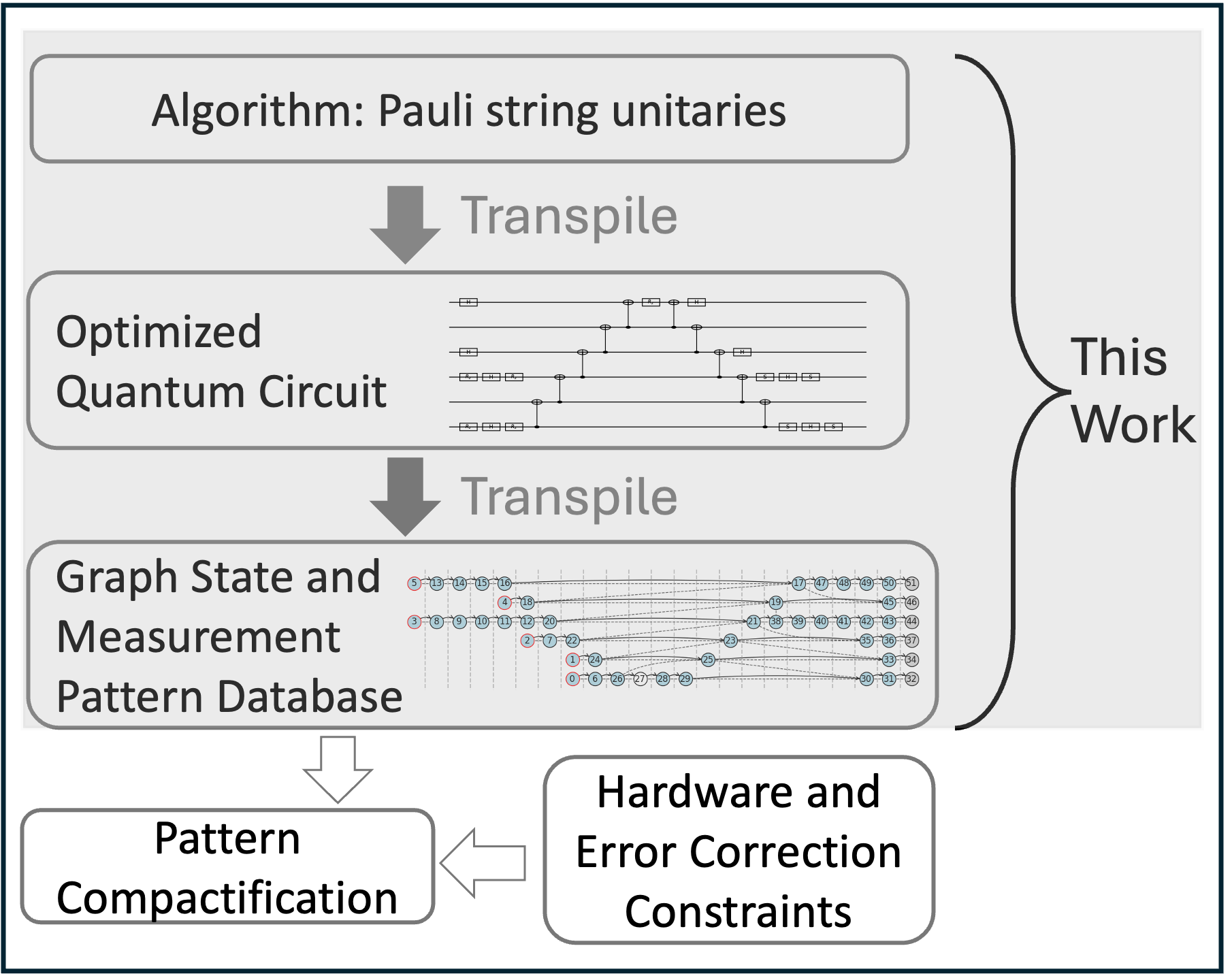}
\caption{Schematic of the workflow used to define v1.0 of the measurement pattern library QPatLib.  We start with a list of Pauli string unitaries used in, e.g., first order Suzuki-Trotter time evolution of a fermionic Hamiltonian. In this case the Hamiltonian is expressed as a sum of Pauli strings and we seek to build a pattern for just one Trotter step (more Trotter steps can be added by concatenating patterns).  The first transpile step requires the strings to be grouped into commuting subsets.  A quantum circuit is then created from the unitary of Pauli strings. The second transpile step builds a measurement pattern on a graph state, one pattern per subset.  The patterns are stored in QPatLib as readable files.  The bottom boxes indicate prospects for pattern compactification based on hardware and error correction needs.    }
\label{fig_workflow_schematic}
\end{figure}

This motivates QPatLib: rather than presenting a single purportedly optimal pattern (or a new compiler), QPatLib records validated measurement patterns together with a minimal set of pattern metrics and multiple commuting-subset strategies \cite{BARTHEL2020,GOKHALE2020,SMITH2025,KALDENBACH2025} as explicit benchmark axes. QPatLib takes a bottom-up approach \cite{KALDENBACH2025a,KALDENBACH2025} in which patterns for Pauli-string unitaries are generated at the subset level so they can be optimized independently and later concatenated, enabling scalable benchmarks across models and constraints. Commuting-set strategies provide a systematic knob for trading off pattern properties, and the library is intended as a standardized testbed and exchange format for evaluating compactification and optimization protocols under differing hardware and error-correction assumptions.

Figure~\ref{fig_workflow_schematic} shows a schematic workflow for the construction and use of v1.0 of the library. Models of interest are encoded into $n_q$ qubits. We start by arranging Pauli strings for these qubits into commuting subsets. The total possible number of Pauli strings is large, $4^{n_q}$. For $n_q<6$, we construct the full set of $4^{n_q}$ Pauli strings to allow implementation of any qubit Hamiltonian expressible as a Pauli expansion. This exhaustive Pauli-string set provides unit tests for MBQC transpilers, optimizers, and hardware (enabling quick correctness and determinism checks) and a labeled training set for machine learning--guided compactification via the stored pattern metrics. For $n_q\geq 6$ we focus on string sets for specific problems. We select Pauli strings used to define electronic structure Hamiltonians of benchmark molecules \cite{SAWAYA2024} (other unitary Pauli-string sets will be added in later versions). We then transpile \cite{JAVADI-ABHARI2024} the unitaries of these Pauli subsets into quantum circuit subsets. Each subset is next transpiled \cite{SUNAMI2022} into a measurement pattern and stored in the library. These patterns can then be concatenated into a useful operator. Here we consider Suzuki--Trotter Hamiltonian time evolution as an example, but the Pauli-string unitaries stored in the library can be used in other algorithms.

\begin{table}[t]
\centering
\begin{tabular}{||c c  ||} 
 \hline
 & List of Pattern Properties   \\ [0.5ex] 
 \hline\hline
 $n$ & Node number  \\ 
 \hline
  $n_e$ & Number of graph edges \\
 \hline 
 $m_d$ & Maximum graph degree \\
 \hline 
 $n_P$ & Number of Pauli measurements \\
 \hline
$m_w$ & Maximum pattern width \\
 \hline
 $n_l$ & Number of measurement layers \\
 \hline
 $m_{ld}$ & Maximum layer distance \\
 \hline
\end{tabular}
\caption{Measurement pattern properties to be optimized and cataloged in the library.  The first three rows indicate properties of the graph state.  The remaining rows are properties that also depend on measurement protocols.}
\label{tab_pattern_properties}
\end{table}

The library can be used to optimize subset patterns prior to concatenation. Criteria are then easier to implement than in a top-down approach because the size of the starting patterns can be controlled to scale up. Table~\ref{tab_pattern_properties} lists parameters that control or define pattern properties. Users will be able to choose a model and strategy for grouping commuting sets of Pauli strings, load a JSONL file from the library, compute their cost function over the Table~\ref{tab_pattern_properties} metrics, and run compactification candidates using, e.g., simulated annealing \cite{KALDENBACH2025a,SHARMA2026}, machine learning \cite{DONG2025,LI2025}, or other methods. The library is open and intended for user write-back with new patterns in the format defined here. A key intended use for the dataset is to serve as a standard for future comparisons of competing compactification algorithms on a dataset that scales to important problems.

QPatLib curates standardized and validated patterns. The patterns are provided in human-readable JSONL format and are paired with the equivalent quantum circuit file in QASM~3.0 format; code for producing the circuits and patterns is included for reproducibility; and Table~\ref{tab_pattern_properties} metrics are included for hardware-aware optimization and comparison. Although these patterns are generated using existing open-source components (for circuit transpilation, commuting-set grouping, and MBQC pattern extraction), the contribution of this work is not a set of transpilation steps. Rather, QPatLib provides a curated, validated, and standardized benchmark dataset and exchange format that enables systematic evaluation of pattern compactification and optimization methods, including user write-back of improved patterns under the same schema and provenance conventions. Finally, v1.0 does not contain patterns produced by a compactification algorithm, making it a baseline dataset for future compactification studies and comparative benchmarking.

The paper is organized as follows. Section~\ref{sec_preliminaries} reviews preliminaries and notation, specifically: the definition of graph states, transpilation of quantum circuits to standard-format measurement patterns, compactification, and the Suzuki--Trotter approximation to time-evolved electronic structure Hamiltonians. Section~\ref{sec_measurement_patterns_from_circuits} presents the central results. Section~\ref{sec_library_of_patterns} details how the library is constructed, and Sec.~\ref{sec_numerical_results} presents results demonstrating the design principle behind the subset approach, including numerical trends for molecular structure models in the library. Section~\ref{sec_compactification_examples} briefly discusses compactification approaches and how they can be applied to the library. Section~\ref{sec_outlook} summarizes and discusses the outlook, including routes for expansion of the library and further use cases. Section~\ref{sec_data_availability} lists resources for code and data availability.  Appendices~\ref{sec_encoded_Be2}, \ref{sec_computational_mehods}, \ref{sec_library_format}, and \ref{sec_validation} detail an example encoded molecular Hamiltonian, computational methods, the data format of the library, and validation methods, respectively.

\section{Preliminaries and Notation}
\label{sec_preliminaries}

\subsection{Graph States} 

Graph states are defined by a set of $n$ nodes $\{\mathcal{N}\}$ and $n_e$ edges $\{ \mathcal{E} \}$ \cite{HEIN2006}. If, for the $i^{\text{th}}$ node, we define the usual qubit Pauli matrices and the Identity as $\hat{X}_i, \hat{Y}_i, \hat{Z}_i$ and $\hat{I}_i$, respectively, then eigenstates of $\hat{X}_i$ with eigenvalues $\pm 1$ are $\vert \pm \rangle_i$. We use this basis to define a graph state as:
\begin{align} 
\prod_{\mathcal{E}_{j,k} \in \{ \mathcal{E}\} }\widehat{CZ}_{j,k} \bigotimes_{\mathcal{N}_i\in \{\mathcal{N}\}} \vert +\rangle_i
\end{align}
where $\widehat{CZ}_{j,k}$ defines the controlled-$Z$ operation between nodes $j$ and $k$.

Graph states will be characterized by properties related to their application in constructing specific MBQS measurement patterns. We will consider only open graph states, i.e., graph states with an input and output for information processing.  Here ‘input nodes’ means nodes that initially carry the input state; these nodes may still be measured during the pattern, whereas ‘output nodes’ are the unmeasured nodes at the end.  The first three rows of Table~\ref{tab_pattern_properties} list graph-state properties. $n$ and $n_e$ are implicit in the definition of the graph state. The maximum degree, $m_d$, denotes the largest number of edges emanating from any single node in the graph. For example, a two-dimensional square-lattice cluster state has $m_d=4$, and a chain (a wire) has $m_d=2$. Graph-state properties should be optimized using hardware and algorithmic constraints.

\subsection{Transpiling Circuits into Measurement Patterns on Graphs}
\label{sec_transpiling}

Quantum circuits are mapped to measurements on graph states using the standard MBQC mapping \cite{RAUSSENDORF2001,RAUSSENDORF2003}. As more quantum gates are applied to the corresponding quantum circuit, the graph is appended with nodes and edges. Single-qubit gates map to measurements on several new nodes along wires, whereas two-qubit gates, e.g., CNOTs, are mapped to measurements on two new nodes and three new edges. Pauli string unitaries typically transpile into circuits with CNOT staircases (CNOT ladders) \cite{Whitfield2011,YORDANOV2020,KALDENBACH2025}. We will see that isolated nearest-neighbor CNOT staircases lead to graphs with degree of at most 4, but circuits with overlapping CNOT staircases can lead to highly non-trivial graphs with higher degree.

The graph state and measurement protocol are defined with the measurement-calculus standard \cite{DANOS2007a}. Reading from left to right, we start with the set of nodes and edges defining the graph state. This is followed by the set of single-qubit projective measurements on all but the output nodes. The projective measurements also include measurement dependencies (signals). The final pattern is accompanied by a set of byproduct Pauli $\hat{X}$ and $\hat{Z}$ operators that correct the output for the randomness of the measurement outcomes. The pattern definition will, in all cases, include signal shifting \cite{BROADBENT2009a}. In this convention, a generic pattern $\mathcal{P}$ can be written \cite{DANOS2007a}: 
\begin{align}
\mathcal{N}_0\cdots \mathcal{N}_{n-1}
\left\{ \mathcal{E} \right\}
\left\{  [\mathcal{M}_i^{\alpha_i}]^{s_i} \right\}
\left\{X^{\{s\}}_l Z^{\{s\}}_l \right\},
\label{eq_pattern_convention}
\end{align}
where $\{[\mathcal{M}_i^{\alpha_i}]^{s_i}\}$ denotes a set of measurements in the $x$--$y$ plane of non-output nodes indexed by $i$ at angle $(-1)^{s_i}\alpha_i$ with respect to the $x$ axis of the Bloch sphere. The measurements can either be Pauli or non-Pauli. Pauli measurements refer to measurements along either the Pauli-$x$ or $y$ directions. $\{X^{\{s\}}_l Z^{\{s\}}_l \}$ denotes a set of correction byproduct operators that act on the output nodes, $l$. $s_i\in\{0,1\}$ is a (possibly composite) signal computed from prior measurement outcomes; it flips the measurement angle $\alpha_i\rightarrow (-1)^{s_i}\alpha_i$.

As an example measurement pattern, consider the identity operation (teleportation) acting on a single-qubit input state. The pattern executes on a 5-node chain graph state with one input at $\mathcal{N}_0$, one output at $\mathcal{N}_4$, and 3 internal nodes \cite{RAUSSENDORF2003}:
\begin{align}
\mathcal{P}^{5-I}&=
\mathcal{N}_0\mathcal{N}_1\mathcal{N}_2 \mathcal{N}_3  \mathcal{N}_4
 \mathcal{E}_{0,1}\mathcal{E}_{1,2} \mathcal{E}_{2,3} \mathcal{E}_{3,4} \nonumber \\ 
 &\cdot\mathcal{M}_0^{0} \mathcal{M}_1^{0} [\mathcal{M}_2^{0}]^{\{1\}}\mathcal{M}_3^{0}
Z_4^{\{0,2\}}
X_4^{\{1,3\}}
\label{eq_p5_identity}
\end{align}
The set of superscripts indicates measurement outcomes from those nodes. For example, $\{0,2\}$ on the $Z$ byproduct indicates the set of signals $s_0$ and $s_2$.

\begin{figure}[t]
\centering
\includegraphics[width=0.46\textwidth]{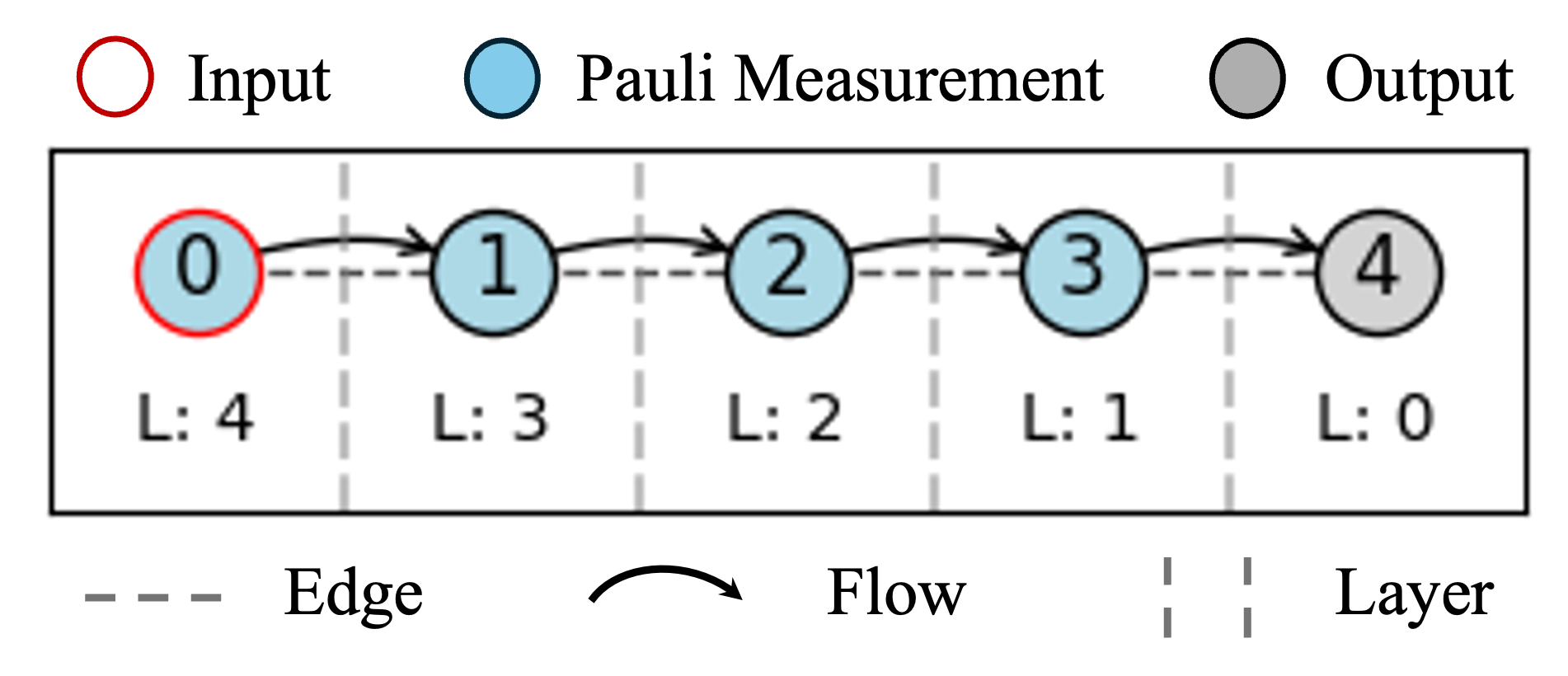}
\caption{A schematic of the measurement pattern for executing the identity operation using a 5-node chain graph, $\mathcal{P}^{5-I}$ [See Eq.~\eqref{eq_p5_identity}].  Circles denote nodes: Inputs (red), Pauli-$x$ or Pauli-$y$ Measurement (blue-filled), and Output (grey-filled).  The numbers in circles index nodes.  The horizontal dashed lines denote graph state edges.  The vertical dashed lines denote measurement layers and are labeled with L.  The solid arrows denotes causal flow found in the graph. The byproduct operators are not shown. }
\label{fig_five_node_identity}
\end{figure}

Figure~\ref{fig_five_node_identity} plots a schematic of $\mathcal{P}^{5-I}$. While the horizontal dashed lines denote edges, the solid lines with arrows denote causal flow \cite{DANOS2006} (as opposed to generalized flow \cite{BROWNE2007,BACKENS2021a} or Pauli flow \cite{SIMMONS2021,MHALLA2025a,MITOSEK2026}). Flow shows that the measurement-layer structure preserves causality. The patterns stored in QPatLib v1.0 are all consistent with causal flow and the measurement-layer structure is chosen in the maximally delayed (i.e., minimal-depth / maximal-parallelism) convention \cite{DANOS2007a,BROWNE2007,MHALLA2008,BROADBENT2009a,PIUS2015}.

The last four rows of Table~\ref{tab_pattern_properties} define measurement-pattern properties. $n_P$ is the number of Pauli measurements. The pattern width, the number of measurements in any single measurement layer, is another important pattern parameter.  Wide patterns leverage parallelism but the pattern width is constrained by hardware.  Table~\ref{tab_pattern_properties} lists the maximum pattern width, $m_w$, corresponding to the largest number of measurements in any single measurement layer.  The number of measurement layers, $n_l$, is an upper bound on the maximum layer distance, $m_{ld}$. $m_{ld}$ is the largest number of layers traversed by any edge in the pattern. $m_{ld}$ is, in particular, a non-trivial hardware requirement since it represents the longest edge that must be maintained as measurements are performed. For Fig.~\ref{fig_five_node_identity}, $n_l=5$ (where measurements on input and output are included in the definition of $n_l$), $n_P=4$, and $m_{ld}=m_w=1$.

\subsection{Compactification}

We define pattern compactification, $\mathbbm{C}$, as a set of operations on a measurement pattern that lowers the node count while leaving the logical unitary intact. Here, $\mathbbm{C}$ is defined to not act on input and output nodes. There are many possible routes to compactification. The most common examples use local complementation on the graph state \cite{HEIN2006} and the corresponding removal of Pauli measurements on a predefined measurement pattern to yield a smaller measurement pattern that is equivalent to the original unitary, up to local Clifford operations. This procedure (single-qubit Pauli removal, updated byproduct operations, and local Clifford decorations) maps to the vertex-minor problem in graph theory \cite{DAHLBERG2018a} and will be denoted $\mathbbm{LC}$. $\mathbbm{LC}$ is an example of $\mathbbm{C}$. Compactification examples are discussed in Sec.~\ref{sec_compactification_examples}.

Given a pattern, many compactified equivalent patterns are possible. As pointed out in Fig.~\ref{fig_workflow_schematic}, we leave specific compactifications of the patterns we produce as separate, use-dependent tasks. Nonetheless, we can characterize the potential for a given pattern to be optimized.  $n_P$, by construction, 
 upper-bounds the maximum number of nodes that could be removed by Pauli-measurement elimination within an $\mathbbm{LC}$-equivalence class, subject to the usual constraints of the measurement calculus.  For example, $n_P=4$ and $n=5$ for $\mathcal{P}^{5-I}$, implying that we can, in principle, remove all but the output node for this case (here $\mathbbm{LC}$ merges the input and output nodes so the input node also persists). This is expected for the identity operation. We note that $n_P$ does not apply as a rigorous bound to $\mathbbm{C}$ in general, but it is a rigorous bound for $\mathbbm{LC}$.

\subsection{Quantum Simulation with Time Evolution in the Suzuki-Trotter Approximation}
\label{sec_quantum_simulation}

We now turn to the quantum-simulation step used to define patterns in v1.0 of QPatLib.  We motivate the need for Pauli string unitaries in the context of Hamiltonian time evolution but Pauli string unitaries are used in many other algorithm families.  To begin, consider an algorithm seeking to time evolve a static Hamiltonian capturing interacting fermions:
\begin{align}
\hat{H}_{F}=\sum_{q_1q_2}h_{q_1q_2}\hat{C}^\dagger_{q_1}\hat{C}^{\vphantom{\dagger}}_{q_2}+\frac{1}{2}\sum_{\{q\}}h_{\{q\}}\hat{C}^\dagger_{q_1}\hat{C}^\dagger_{q_2}\hat{C}^{\vphantom{\dagger}}_{q_3}\hat{C}^{\vphantom{\dagger}}_{q_4},
\label{eq_fermion_model}
\end{align}
where the second-quantized operator $\hat{C}^\dagger_{q}$ creates a fermion in an orbital $q$. $q$ includes the spin degree of freedom and $\{q\}=q_1,q_2,q_3,q_4$. $h_{q_1q_2}$ and $h_{q_1q_2q_3q_4}$ are single- and two-particle matrix elements, respectively, that, for the molecules used here, capture electronic structure. For $n_q<6$, the patterns in QPatLib can be used for any Hamiltonians captured by Eq.~\eqref{eq_fermion_model}. QPatLib labels these patterns used for $n_q<6$ as ``arbitrary'' models. For $n_q\geq 6$, v1.0 focuses on the electronic structure of active orbitals in benchmark diatomic molecules as defined in the HamLib library \cite{SAWAYA2024}: \ce{Be2}, \ce{H2}, \ce{BH}, \ce{OH}, \ce{NH}, \ce{C2}, \ce{N2}, and \ce{Na2}.

Interacting models of fermions map to qubit-encoded models using the Jordan--Wigner transformation \cite{JORDAN1928b} such that, for any model, the fermionic operators map to linear combinations of $n_s$ Pauli strings, $\hat{P}_s$.  The fermion Hamiltonian then becomes the qubit-encoded Hamiltonian:
\begin{align}
\hat{H}=\sum_{s=0}^{n_s-1}c_s \hat{P}_s,
\label{eq_Hamiltonian_string}
\end{align}
where the number of encoded qubits corresponds to twice the number of active molecular orbitals (the factor of 2 arising from the electron spin in the model). The coefficients $c_s$ are determined by $h_{q_1q_2}$ and $h_{q_1q_2q_3q_4}$ for a given model \cite{SAWAYA2024}. Appendix~\ref{sec_encoded_Be2} shows an example encoding for \ce{Be2}, which needs 6 qubits to encode the active electrons.

A central aim of quantum simulation algorithms in the phase estimation family \cite{Kitaev1995,Abrams1999,Ortiz2001,Somma2002,ASPURU-GUZIK2005a,Poulin2018} and related hybrid algorithms \cite{MOTTA2020,SHEN2023a,YU2025a,BHARTI2021,HAUG2022,LEE2024,ZHAO2025,LEE2025a} is to evaluate the propagator on a quantum register:
\begin{align}
e^{-i\hat{H}t} \vert \psi_0\rangle
\label{eq_propagator}
\end{align}
where $\vert \psi_0\rangle$ denotes an initial state and $t$ is time. We assume the first-order Suzuki--Trotter \cite{Trotter1959,Suzuki1976} decomposition with a cutoff in the number of Trotter steps, $p_c$: 
\begin{align}
e^{-i\hat{H}t}= \left[
\prod_{s\in J}e^{-i \delta t c_{\pi(s)}\hat{P}_{\pi(s)}}
\right]^{p_c} 
+\epsilon(p_c,t),
\end{align}
where $\delta t\equiv t/p_c$ and $\epsilon(p_c,t)$ is the Suzuki--Trotter truncation error. $\pi$ denotes one of the $n_s!$ ordering choices for the Pauli strings such that $\pi(s)$ is an element of the symmetric group. $J$ denotes a finite set of strings. Changing the definition of $\pi$ reorders the product strings.

  Tight bounds on $\epsilon(p_c,t)$ are known \cite{Childs2021} and can be evaluated given the Hamiltonian coefficients and set of Pauli strings. To see the trend, note that the largest contribution to $\epsilon(p_c,t)$ is of order:
\begin{align}
 \mathcal{O}\left[ \frac{\vert\vert c_{s'}c_{s''} [\hat{P}_{\pi(s')},\hat{P}_{\pi(s'')}]\vert\vert t^2}{p_c}\right],
\end{align}
where $\vert\vert \cdots \vert\vert$ denotes the spectral norm and $\vert\vert c_{s'}c_{s''} [\hat{P}_{\pi(s')},\hat{P}_{\pi(s'')}]\vert\vert$ is the largest norm defined by the pair of Pauli strings indexed by $\pi(s')$ and $\pi(s'')$.

For use of patterns in a Trotter time step, we absorb the time step into the coefficients by redefining the coefficient $\delta t c_s \rightarrow c_s$.  This is equivalent to setting $\delta t$ to unity. The library stores patterns symbolically and sets $\delta t$ to unity.  (When quoting the Hamiltonian in Appendix~\ref{sec_encoded_Be2}, $c_s$ denotes the physical coefficient before multiplying by $\delta t$.)  The replacement allows pattern construction for just one Trotter step and $p_c>1$ follows by concatenation. For example, consider a Hamiltonian with 4 Pauli strings and an ordering choice $\pi=\{2,0,1,3\}$. A single Trotter (unit time) step would require the operator $e^{-ic_2 \hat{P}_2}e^{-ic_0 \hat{P}_0}e^{-ic_1 \hat{P}_1}e^{-ic_3 \hat{P}_3}$. Concatenating this operator $p_c$ times on an initial state approximates Eq.~\eqref{eq_propagator}. The central aim is then to construct a protocol for evaluating all concatenated elements in the propagator, i.e., all $e^{-i c_{\pi(s)}\hat{P}_{\pi(s)}}$ or combinations thereof, for a single Trotter step. Changing the choice of $\pi$ leads to different unitaries, but the difference vanishes as $\epsilon(p_c,t)\rightarrow0$.

\section{Measurement Patterns from Quantum Circuits}
\label{sec_measurement_patterns_from_circuits}

In this section we use the background from the prior sections to define measurement patterns for Pauli string unitaries. Sec.~\ref{sec_library_of_patterns} defines the protocol used to establish patterns in QPatLib.  Sec.~\ref{sec_numerical_results} then presents results detailing the subset design principle. Numerical results obtained from analyzing large collections of measurement patterns illustrate this design principle through empirical trends for the Pauli string unitaries defining the models of the chosen molecules.

\subsection{Library of Patterns for Evolution of Subsets of Commuting Pauli Strings}
\label{sec_library_of_patterns}

\begin{figure*}[t]
\centering
\includegraphics[width=\textwidth]{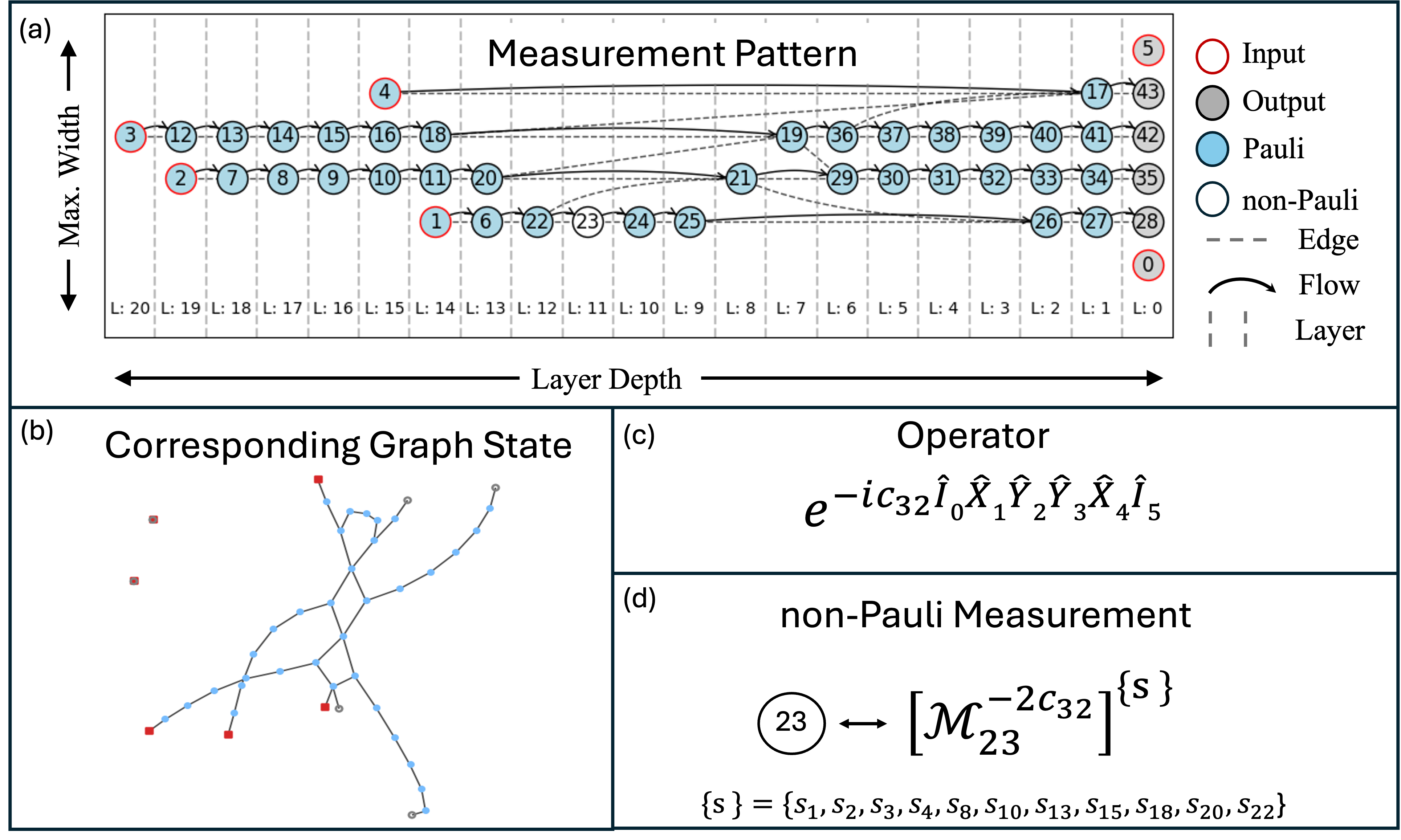}
\caption{(a) An example measurement pattern that executes the unitary $e^{-i  c_{32}\hat{P}_{32}}$ on the state prepared on the input nodes (0-5), where  
$\hat{P}_{32}=\hat{I}_0\hat{X}_1\hat{Y}_2\hat{Y}_3\hat{X}_4\hat{I}_5$.  This pattern represents the $32^{\text{nd}}$ subset of the O-O strategy for \ce{Be2} (See Table~\ref{tab:be2-pauli-strings}). The empty circle at node 23 represents a non-Pauli measurement.  The graph state edge $\mathcal{E}_{4,17}$ spans 14 measurement layers to establish the longest for this subset.  Output nodes are 0, 28, 35, 42, 43 and 5. (b) A schematic of the nodes and edges in the corresponding graph state where the inputs are red.  (c) The operator executed by the measurement pattern acting on the encoded qubits 0-5.  (d) The non-Pauli measurement at node 23 showing that the measurement angle depends on the Hamiltonian coefficient $c_{32}$ as well the set of signal domains $\{s\}$ required from prior measurements.  
}
\label{fig_example_pattern_32_IXYYXI}
\end{figure*}

This section defines measurement patterns using the workflow from Fig.~\ref{fig_workflow_schematic}: (i) we start with a set of strings and coefficients; (ii) we arrange collections of strings into commuting subsets; (iii) each subset is then used to build a single measurement pattern; and (iv) the pattern from each subset is validated and stored in the library.

We start by defining commuting subsets of Pauli strings. Given a collection of Pauli strings of the same length (including the identity operator), we sort the strings into $n_{\text{ss}}$ subsets such that the Pauli strings within each subset commute with each other but not necessarily between subsets. If there are $n_{\text{ss}}^{(\ell)}$ strings in the $\ell^{\text{th}}$ subset, then $n_{\text{s}}=\sum_{\ell=0}^{n_{\text{ss}}-1}n_{\text{ss}}^{(\ell)}$.

There are many choices for subset definitions. In v1.0 we choose three example subset definitions to enable relative comparisons between the resulting measurement patterns. We start with the \emph{One-to-One} (O-O) strategy, where there is only one Pauli string per subset such that $n_{\text{ss}}=n_s$ (Appendix~\ref{sec_encoded_Be2} details the example of \ce{Be2}, where each row of Table~\ref{tab:be2-pauli-strings} is a subset with the O-O strategy). Each subset (each string in the O-O case) is then converted to an evolution unitary, $\exp{[-ic_{\pi(s)}\hat{P}_{\pi(s)}]}$, where the ordering $\pi$ is assumed.

Following the workflow of Fig.~\ref{fig_workflow_schematic}, the evolution unitary is converted into a quantum circuit. Evolution unitaries of a single Pauli string lead to quantum circuits containing one pair of CNOT staircases per unitary for the O-O strategy. We then use the quantum circuit to construct a measurement pattern as discussed in Sec.~\ref{sec_transpiling}.

Figure~\ref{fig_example_pattern_32_IXYYXI} shows an example pattern for the exponential of the Pauli string $\hat{I}_0\hat{X}_1\hat{Y}_2\hat{Y}_3\hat{X}_4\hat{I}_5$, the $32^{\text{nd}}$ string for \ce{Be2} from Appendix~\ref{sec_encoded_Be2}. Panel (a) shows the measurement pattern that executes $e^{-i  c_{32}\hat{P}_{32}}$ on inputs defined on qubits $0$--$5$ (red circles). Note that all but the output nodes are measured (filled grey circles). The output is to be used in concatenation with the next pattern. Panels (b) and (c) show the corresponding graph state and operator, respectively. Panel (d) shows that the open circle at node 23 corresponds to a non-Pauli measurement in the $x$--$y$ plane at an angle determined by $c_{32}$. Here, the set of feedforward integers $\{s\}$ is also displayed. 

We can use Fig.~\ref{fig_example_pattern_32_IXYYXI} to read off the pattern parameters listed in Table~\ref{tab_pattern_properties}. We find $n=44$ and $n_P=37$. From panel (b) it is easiest to see $m_d=4$ and $n_e=45$. For this pattern, causal flow is consistent with $n_l=21$ and $m_w=6$. From panel (a) we also find $m_{ld}=14$ because the edge between nodes 4 and 17 traverses 14 layers.

Once measurement patterns are defined for each subset, we can concatenate each subset to construct a Trotter step. The pattern for a single Trotter step is:
\begin{align}
\mathcal{P}^{\text{O-O}}=\mathcal{P}^{\text{O-O}}_{\pi(0)} \cup \mathcal{P}^{\text{O-O}}_{\pi(1)} \cup \cdots \cup \mathcal{P}^{\text{O-O}}_{\pi(n_{\text{s}}-1)}, 
\end{align}
where $\mathcal{P}^{\text{O-O}}_{\pi(s)}$ is the measurement pattern corresponding to the evolution of the $s^{\text{th}}$ string under the ordering $\pi$, $e^{-ic_{\pi(s)} \hat{P}_{\pi(s)}}$, and $\cup$ implies concatenation of patterns such that the output of the first pattern replaces the input of the second pattern, the output of the second replaces the input of the third pattern, and so on. Section~\ref{sec_compactification_examples} discusses opportunities for compactification using the subset approach.

The commuting-subset approach allows the choice of new strategies to grow subset size beyond the O-O strategy. As a second example, we consider the \emph{Smallest-to-Last} (S-L) strategy \cite{HAGBERG2008}. The problem of defining collections of commuting Pauli strings is mapped to a network problem \cite{GOKHALE2020} such that vertices are Pauli strings, links represent non-commutation, and colors correspond to commuting subsets (note that the network of vertices for defining subsets is distinct from the graph state). The S-L ordering heuristic \cite{HAGBERG2008} maintains current vertex degrees of the network via a bucket queue and repeatedly removes a minimum-degree vertex while updating its neighbors' degrees. It then colors vertices in the reverse of this removal order. The resulting Pauli-string subsets greatly diminish in size as more subsets are created.

\begin{table}[ht]
\centering
\begin{tabular}{||c | c | c ||} 
 \hline
Molecule & Smallest-to-Last & One-to-One  \\ [0.5ex] 
 \hline\hline
 \ce{Be2} & 7 & 62  \\ 
 \hline
 \ce{H2} & 12 & 185  \\
 \hline
 \ce{BH} & 20 & 276  \\
 \hline
 \ce{OH} & 35 & 631  \\
 \hline
 \ce{NH} & 58 & 1086  \\ 
  \hline
 \ce{C2} & 82 & 1884  \\ 
  \hline
 \ce{N2} & 153 & 4578  \\ 
  \hline
 \ce{Na2} & 164 & 6509  \\ 
 \hline
\end{tabular}
\caption{Right column: Number of Pauli strings used to define the encoded Hamiltonian for each molecule (the O-O subset strategy).  Middle column: Number of subsets of commuting Pauli strings using the S-L strategy.}
\label{tab_subsets}
\end{table}

The middle column of Table~\ref{tab_subsets} shows the number of subsets found for each molecule using the S-L strategy to group their corresponding Pauli strings. Here we have $n_s > n_{\text{ss}} > 1$. The right column of Table~\ref{tab_subsets} shows $n_s$ for the O-O strategy for comparison (Table~\ref{tab:be2-subsets} in Appendix~\ref{sec_encoded_Be2} shows the 7 S-L subsets for \ce{Be2} as an example). The products of Pauli-string evolution unitaries using the S-L strategy are equivalent to the unitaries built with the O-O strategy if the string order is the same.

Using the S-L strategy to build patterns, we can again follow the workflow shown in Fig.~\ref{fig_workflow_schematic}. We transpile the circuit derived from the strings in each subset to obtain one pattern per subset, $\mathcal{P}^{\text{S-L}}_{\ell}$, where $\ell$ indexes the pattern obtained from the $\ell^{\text{th}}$ subset of strings. The total pattern is then obtained by concatenating the patterns obtained from each subset:
\begin{align}
\mathcal{P}^{\text{S-L}}=\mathcal{P}^{\text{S-L}}_0 \cup \mathcal{P}^{\text{S-L}}_1 \cup \cdots \cup \mathcal{P}^{\text{S-L}}_{n_{\text{ss}}-1},
\end{align}
where, as we see from Table~\ref{tab_subsets}, many fewer concatenations are needed. However, as we will see in Sec.~\ref{sec_numerical_results}, the circuit transpilation step in Fig.~\ref{fig_workflow_schematic} leads to significant trade-offs between the O-O and S-L strategies because the circuit CNOT staircases in the O-O strategy are kept apart (by design), whereas circuit optimization tends to overlap CNOT staircases in the S-L strategy. In the following, we refer to increases in overlapping staircases as increasing circuit ``complexity''.  We will see that circuit complexity leads to graph states and patterns with more demanding hardware requirements.

The third and final subset strategy, the \emph{Full} strategy, targets a limit opposing the O-O strategy. The Full strategy uses only a single subset (for one Trotter step). The circuit generated from evolution of the full Hamiltonian (such that $J_{\text{full}}$ denotes all strings):
\begin{align}
\prod_{s\in J_{\text{full}}} e^{-i c_{\pi(s)}\hat{P}_{\pi(s)}},
\end{align}
is then transpiled into a single large pattern, $\mathcal{P}^{\text{Full}}$, with $n_{\text{ss}}=1$.

To summarize this section, we established three strategies for building patterns from evolution unitaries of subsets of commuting Paulis to reach three extremes. We defined the O-O strategy to give the largest number of subsets, $n_s=n_{\text{ss}}$. We then defined the S-L strategy to yield an intermediate case, $n_s > n_{\text{ss}} > 1$. Finally, we defined the Full strategy to give the smallest number of subsets, with $n_s > n_{\text{ss}}=1$. The resulting patterns are $\mathcal{P}^{\text{O-O}}$, $\mathcal{P}^{\text{S-L}}$, and $\mathcal{P}^{\text{Full}}$, respectively.

The central result of this work is the creation of QPatLib, datasets containing the patterns for each subset and the associated quantum circuits for each subset.  Appendix~\ref{sec_computational_mehods} lists computational methods used to produce the library.  The pattern data format is detailed in Appendix~\ref{sec_library_format} where an example library entry for the pattern of Fig.~\ref{fig_example_pattern_32_IXYYXI} is shown.  The pattern validation and check protocols are explained in Appendix~\ref{sec_validation}. The outlook for compactification is discussed in Sec.~\ref{sec_compactification_examples}. The following section details numerical pattern analysis to show that the subset-strategy choices offer distinct trade-offs for compactification and satisfaction of pattern constraints.

\subsection{Design Principles and Empirical Trends}
\label{sec_numerical_results}

This section uses the library to show that subset granularity is a controllable knob that interpolates between low-degree/short-edge patterns and fewer non-Pauli measurements (and thus larger potential $\mathbbm{LC}$ gains), providing trade-offs for MBQS. The trade-offs are discussed in the context of pattern properties listed in Table~\ref{tab_pattern_properties} for the benchmark molecules. We study patterns for a single Trotter step for the benchmark molecules. We use the same string ordering ($\pi$) but different subset strategies (O-O, S-L, and Full). The same string ordering implies that the unitaries generated by the concatenated patterns are the same for each strategy; yet, as we will see, the pattern properties are very different.

We start with the simplest pattern parameter, the maximum width. All measurement patterns in v1.0 of the library have $m_w$ equal to the number of input/output nodes in the graph. In transpiling patterns from circuits, for every input qubit there is a teleportation wire with local operations. When extracting a pattern consistent with flow, the resulting parallel measurement layers have one measurement per wire per round. Hence, $m_w$ equals the number of logical input/output qubits. Compactification can change this and can grow $m_w$ beyond the number of logical input/output qubits.

\begin{figure}[t]
\centering 
\includegraphics[width=0.45\textwidth]{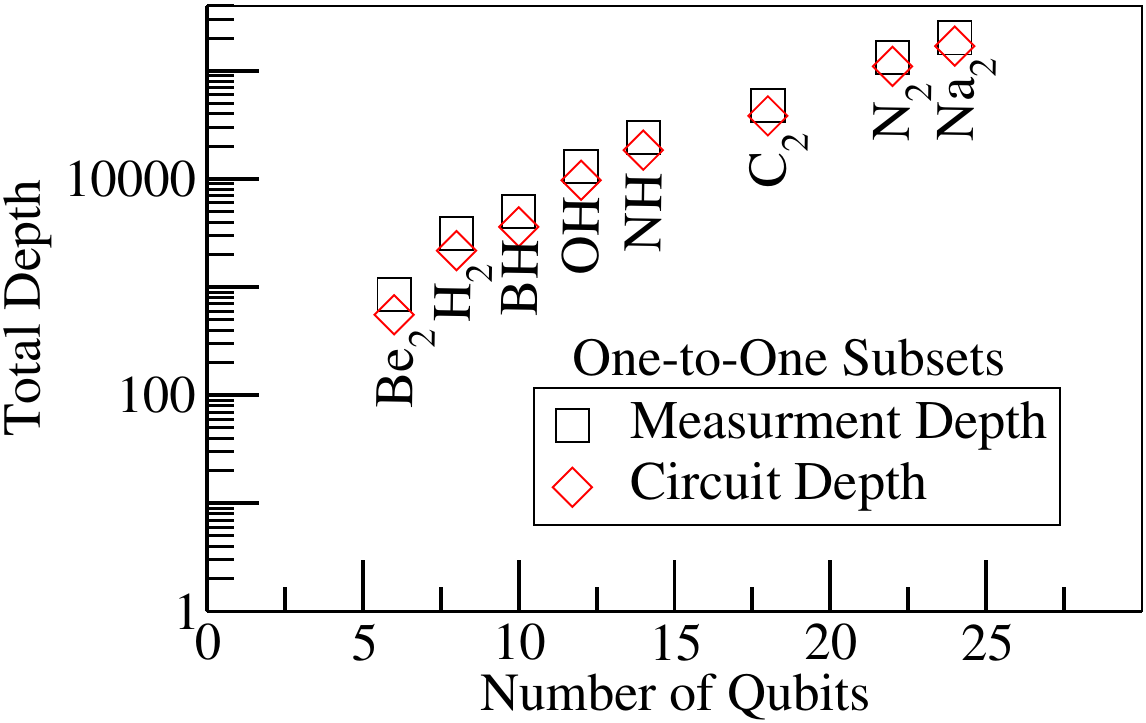}
\caption{Comparison of total depth for the circuit and measurement-based approaches versus the number of qubits encoding each molecule.   Diamonds: The circuits are obtained by transpiling $e^{-ic_{\pi(s)}\hat{P}_{\pi(s)}}$ for each Pauli string and concatenating each circuit (The One-to-One subset strategy).  Circuit depth refers to the number of layers of parallel circuits.  Squares:  Each subset circuit is transpiled into a measurement pattern on a graph state and concatenated.  The measurement depth refers to the number of layers of parallel measurements. 
}
\label{fig_depth_vs_qubits}
\end{figure}

We now turn to depth. Figure~\ref{fig_depth_vs_qubits} compares depths for the benchmark molecules. The diamonds plot the total concatenated circuit depth needed to implement one Trotter step for the O-O strategy. We generated one circuit for $e^{-i  c_{\pi(s)}\hat{P}_{\pi(s)}}$ for each $s$. The circuits were then concatenated with no further optimization. The squares plot the measurement depth, $n_l$, needed to implement $\mathcal{P}^{\text{O-O}}$. We see that the measurement-pattern depth and the circuit depth are of the same order.

\begin{figure}[t]
\centering 
\vspace{-2cm}
\includegraphics[width=0.45\textwidth]{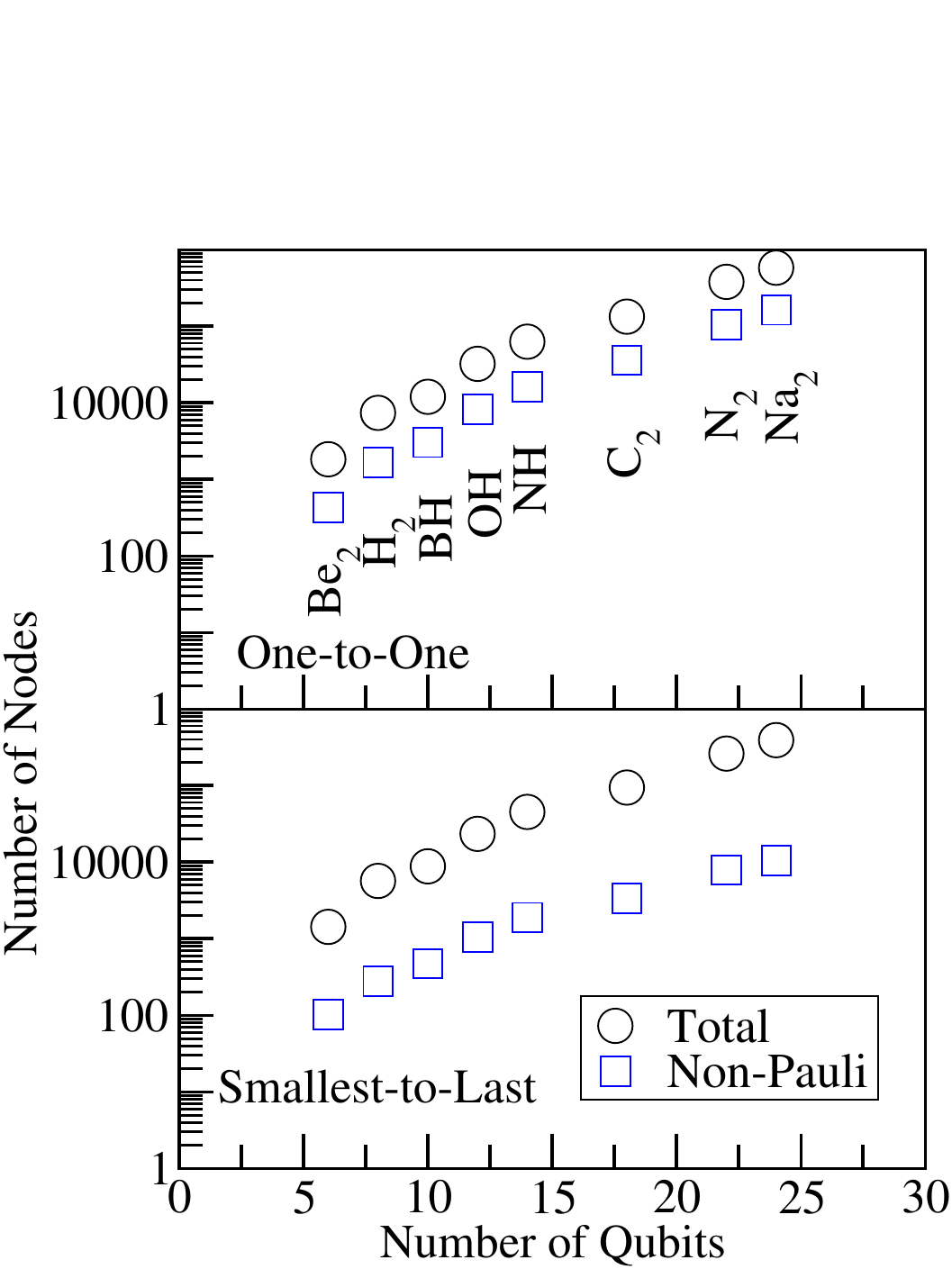}
\caption{Top: The circles plot the total number of nodes in the concatenated set of patterns defining one Trotter step using the One-to-One strategy ($\mathcal{P}^{\text{O-O}}$) for each benchmark molecule.  The squares plot the number of non-Pauli measurements.  The vertical gaps between the squares and circles show possible $\mathbbm{LC}$ gains available by removing Pauli measurements (excluding output nodes).  Bottom: The same as the top panel but for the Smallest-to-Last strategy defining $\mathcal{P}^{\text{S-L}}$.
}
\label{fig_measurement_count_vs_qubit_number}
\end{figure}

We now turn to the total node counts. The top panel of Fig.~\ref{fig_measurement_count_vs_qubit_number} shows the total number of nodes (circles), and the number of non-Pauli measurements (squares) needed to build $\mathcal{P}^{\text{O-O}}$. The vertical gap between the circles and squares is $n_P$ (excluding the output nodes). The gap between circles and squares shows the best possible gain from $\mathbbm{LC}$ on $\mathcal{P}^{\text{O-O}}$. Here we see that $\mathbbm{LC}$ can, at best, return an order-of-magnitude gain in measurement count for the O-O strategy.

The bottom panel of Fig.~\ref{fig_measurement_count_vs_qubit_number} shows the same as the top panel, but for $\mathcal{P}^{\text{S-L}}$. Here we see that the total number of nodes is nearly the same, but the possible $\mathbbm{LC}$ gains (the gaps between circles and squares) are as large as two orders of magnitude for some of the molecules, much better than the O-O strategy. The better (possible) gains for the S-L strategy arise because circuit-level optimization acting on the larger circuit reduced the number of non-Clifford gates by simply combining gates (see Appendix~\ref{sec_computational_mehods}). This reduction, in turn, lowered the number of non-Pauli measurements in $\mathcal{P}^{\text{S-L}}$ compared to $\mathcal{P}^{\text{O-O}}$. The trend we find is that larger subsets require fewer non-Pauli measurements due to circuit-level rewrites.

We find that the possible $\mathbbm{LC}$ gains in the S-L subset strategy come with costs in other pattern parameters. The maximum degree of the underlying graph state is a key hardware constraint. The O-O strategy leads to patterns with a maximum degree of 4 for all of the benchmark molecules studied (see, e.g., Fig.~\ref{fig_example_pattern_32_IXYYXI}b). This is due to the isolation of CNOT staircases in the defining circuit. Table~\ref{tab_max_degree} shows the maximum degree for the S-L strategy. Here we see an increase in the maximum degree as we consider molecules with an increasing number of active orbitals. Increasing circuit complexity increases $m_d$.

\begin{table}[h]
\centering
\begin{tabular}{||c | c  ||} 
 \hline
Molecule & Max. degree  \\ [0.5ex] 
 \hline\hline
 \ce{Be2} & 11   \\ 
 \hline
 \ce{H2} & 10   \\
 \hline
 \ce{BH} & 14  \\
 \hline
 \ce{OH} & 14   \\
   \hline
 \ce{NH} & 18   \\ 
  \hline
 \ce{C2} & 14   \\ 
  \hline
 \ce{N2} & 28   \\ 
  \hline
  \ce{Na2} & 34   \\ 
  \hline
\end{tabular}
\caption{$m_d$ found for the graph states in $\mathcal{P}^{\text{S-L}}$ for each molecule.}
\label{tab_max_degree}
\end{table}

Maximum layer distance is another important hardware constraint. The edge traversing the largest number of layers can also be extracted directly from the pattern library. We find that subset strategies with larger subsets tend to have larger layer distances. The growth arises from increasing circuit complexity.

\begin{table}[t]
\centering
\begin{tabular}{||c || c | c || c | c ||}
 \hline
  & \multicolumn{2}{c||}{O-O} & \multicolumn{2}{c||}{S-L} \\
 \hline
 Molecule & $m_{ld}$ & max($n_l$) & $m_{ld}$ & max($n_l$) \\
 \hline \hline
 \ce{Be2} & 18  & 25  & 143  & 153  \\
 \hline
 \ce{H2}  & 22  & 29  & 265  & 498  \\
 \hline
 \ce{BH}  & 26  & 33  & 411  & 707  \\
 \hline
 \ce{OH}  & 30  & 37  & 735  & 1172 \\
 \hline
 \ce{NH}  & 32  & 39  & 694  & 890  \\
 \hline
 \ce{C2}  & 42  & 49  & 1495 & 1704 \\
 \hline
 \ce{N2}  & 46  & 55  & 3505 & 3810 \\
 \hline
 \ce{Na2} & 50  & 59  & 4920 & 6110 \\
 \hline
\end{tabular}
\caption{Data resulting from a scan over all subsets in $\mathcal{P}^{\text{O-O}}$ (left) and $\mathcal{P}^{\text{S-L}}$ (right). The maximum layer distance and the layer depth of the subset with the largest layer distance, max($n_l$),  are listed.  The comparison shows that max($n_l$) is a close upper bound to $m_{ld}$.}
\label{table_compare_mld_nl}
\end{table}

To study the maximum layer distance, we first show empirically that the number of measurement layers is a good proxy for the maximum layer distance. Table~\ref{table_compare_mld_nl} compares $m_{ld}$ and the maximum $n_{ss}^{(\ell)}$ resulting from a search for the maximum edge-layer span in any subset using the O-O (middle columns) and S-L strategies (right columns). The number of layers in the subset pattern is then recorded as max($n_l$). The search is performed separately for each molecule. We find that $m_{ld}$ is no more than a factor of 2 smaller than max($n_l$), showing that the number of layers in a subset is a good proxy for the maximum edge-layer span in that subset.

\begin{figure}[t]
\centering 
\includegraphics[width=0.47\textwidth]{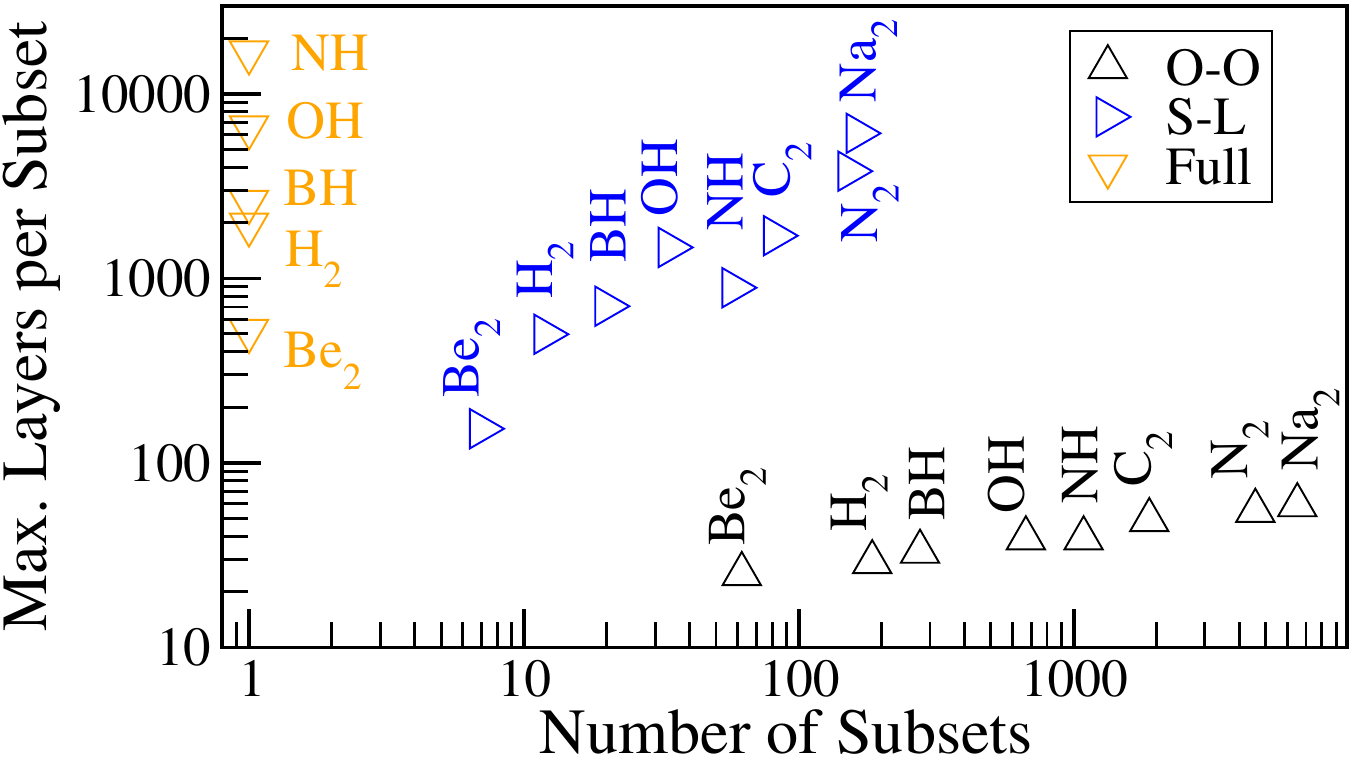}
\caption{The maximum number of measurement layers in any subset versus the number of subsets for three different subset strategies: O-O (up triangle), S-L (right triangle), and Full (down triangle).  The data are taken from measurement patterns for each of the benchmark molecules.  The $y$-axis is a close upper bound on $m_{ld}$ (See Table~\ref{table_compare_mld_nl}). The trends imply that the choice of subset strategy aids in satisfying constraints.  
}
\label{fig_max_layers_vs_subsets}
\end{figure}

We demonstrate an important trend in subset strategies regarding the maximum layer distance by using the maximum number of layers as a proxy. Figure~\ref{fig_max_layers_vs_subsets} plots max($n_l$) versus the number of subsets. The Full strategy has only one subset, but max($n_l$) grows rapidly (approximately exponentially over the studied range) with increasing numbers of encoded qubits. The O-O strategy, by contrast, has slow growth of max($n_l$), with at most 59 layers needed for the most complex molecule, \ce{Na2}. Yet, for the O-O strategy, the number of subsets grows rapidly with the number of encoded qubits. The S-L strategy is intermediate. We see that strategies with a diminishing number of subsets tend to require a larger number of layers per subset (and therefore larger $m_{ld}$). These trends justify an important design principle: we can tune $m_{ld}$ by varying the subset strategy. The subset strategy can therefore be a useful tool to manage compactification schemes that must, in turn, accommodate hardware or other constraints.

We summarize this section by listing design principles displayed in the measurement-pattern data for the benchmark molecules. (i) Maximum pattern width: the number of qubits needed to encode the selected active orbitals fixes $m_w$ regardless of subset strategy. (ii) Depth:  $n_l$ is of the same order as the corresponding circuit depth; both measurement and circuit depths grow rapidly with the number of encoded qubits. (iii) Measurement counts: $n$ increases approximately exponentially over the studied range with the number of encoded qubits; $n_P$ is much larger for subset strategies with large patterns, thus offering the possibility of more $\mathbbm{LC}$ gains with coarser subset partitioning. (iv) Graph-state degree: $m_d$ is fixed to 4 for the O-O subset strategy but increases substantially for the S-L strategy. (v) Maximum layer distance: $m_{ld}$ is well approximated by the number of layers for that subset, max($n_l$). For the O-O strategy, we find that max($n_l$) is bounded but the number of subsets grows rapidly with the number of encoded qubits. The Full subset strategy shows opposite behavior since its single subset has max($n_l$) growing rapidly with the number of encoded qubits. The S-L subset strategy is intermediate.

\section{Compactification Opportunities and Examples}
\label{sec_compactification_examples}

The bottom-up subset+concatenation approach to pattern construction has an important application. As pointed out in the introduction, compactification of large graphs (a top-down approach) is non-trivial. It is especially non-trivial to simultaneously ensure pattern repeatability while also optimizing the parameters in Table~\ref{tab_pattern_properties}.

Compactification can be implemented on the subset patterns. For $\mathcal{P}^{\text{O-O}}$, we can construct:
\begin{align}
\mathbbm{C}[\mathcal{P}^{\text{O-O}}_{\pi(0)}] \cup \mathbbm{C}[\mathcal{P}^{\text{O-O}}_{\pi(1)}] \cup \cdots \cup \mathbbm{C}[\mathcal{P}^{\text{O-O}}_{\pi(n_{\text{s}}-1)} ],
\end{align}
where we see that each subset pattern can be compactified individually, thus avoiding the need to compactify the entire pattern. If hardware constraints permit, we may increase the number of nodes in commuting subsets. In our example, $\mathcal{P}^{\text{S-L}}$ has more nodes per subset. We may then compactify the pattern subsets in the S-L strategy:
\begin{align}
\mathbbm{C}[\mathcal{P}^{\text{S-L}}_1] \cup 
\mathbbm{C}[\mathcal{P}^{\text{S-L}}_2] \cup \cdots \cup \mathbbm{C}[\mathcal{P}^{\text{S-L}}_{n_{\text{ss}}} ],
\end{align}
where fewer pattern concatenations are needed, but compactification will be more non-trivial because it requires operations on larger patterns.

We briefly mention specific compactification approaches. We start with $\mathbbm{LC}$ pair contraction to shorten wires \cite{RAUSSENDORF2003}. QPatLib patterns contain long wires. While these wires might be useful in certain native hardware implementations \cite{LINDNER2009,SCHWARTZ2016,ISTRATI2020,LI2023b,LI2025}, they can be shortened. Consider the 5-qubit identity pattern in Eq.~\eqref{eq_p5_identity}. Performing the removal of a pair of Pauli-$x$ measurements on an isolated wire, $\mathbbm{LC}[\mathcal{P}^{5-I}]\rightarrow \mathcal{P}^{3-I}$, leads to the same unitary, where:
\begin{align}
\mathcal{P}^{3-I}=
 \mathcal{N}_0\mathcal{N}_1\mathcal{N}_2  
 \mathcal{E}_{0,1}\mathcal{E}_{1,2} 
\mathcal{M}_0^{0} \mathcal{M}_1^{0} 
X_2^{\{1\}}
Z_2^{\{0\}},
\end{align}
has 2 fewer Pauli measurements than $\mathcal{P}^{5-I}$. From this example we see that wires in patterns described here, e.g., the pattern in Fig.~\ref{fig_example_pattern_32_IXYYXI}, can be significantly reduced in size or even removed entirely.

\begin{figure}[t]
\centering 
\includegraphics[width=0.35\textwidth]{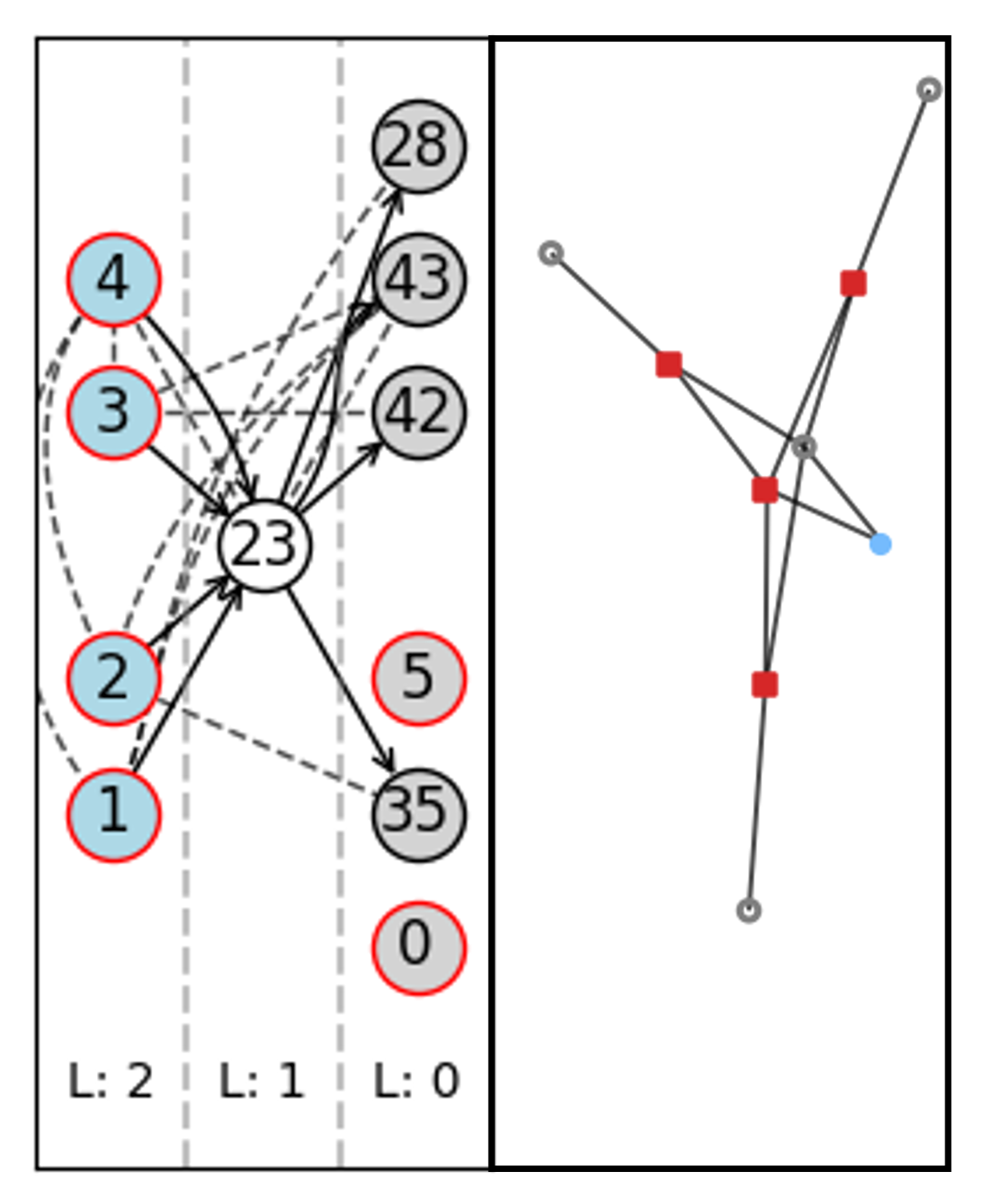}
\caption{
Left: A measurement pattern that produces the same unitary as the pattern in Fig.~\ref{fig_example_pattern_32_IXYYXI} when acting on the input state with $ \vert +\rangle$ on nodes 0-5.   Here the Pauli measurements have been removed from the pattern in Fig.~\ref{fig_example_pattern_32_IXYYXI}. The solid arrows are consistent with generalized flow \cite{BROWNE2007,BACKENS2021a}.  The maximum degree (4) and the maximum width remains the same  whereas  number of the measurement layers is reduced by an order of magnitude. Right: The underlying graph state.
}
\label{fig_example_pattern_32_IXYYXI_compactified}
\end{figure}

Figure~\ref{fig_example_pattern_32_IXYYXI_compactified} shows another example of $\mathbbm{LC}$. We took the example pattern from Fig.~\ref{fig_example_pattern_32_IXYYXI}, $\mathcal{P}^{O-O}_{32}$, and removed internal Pauli measurements \cite{SUNAMI2022}. The resulting pattern plotted in Fig.~\ref{fig_example_pattern_32_IXYYXI_compactified} leaves the critical non-Pauli measurement in the center as well as all input and output nodes but returns the same unitary as $\mathcal{P}^{O-O}_{32}$ when acting on the state with $\vert +\rangle$ on all input qubits.

The example in Fig.~\ref{fig_example_pattern_32_IXYYXI_compactified} happens to optimize all parameters in Table~\ref{tab_pattern_properties}. But, in general, compactification can come with hardware-design costs. For example, the remaining non-Pauli measurements often lead to increases in $m_d$, and restrictions on determinism can increase $m_{w}$.

Examples of $\mathbbm{C}$ that are not examples of $\mathbbm{LC}$ include rewrites. For example, consider an MBQC pattern that executes a concatenated pair of rotations along a wire \cite{RAUSSENDORF2003}. If the first rotation executes a $\pi/4$ rotation about $y$ and the second executes a $-\pi/4$ rotation about $y$, we would, after concatenation, need 6 non-Pauli measurements along the wire. But since this is the identity operation, a simple rewrite shows that we do not need any measurements. As another rewrite example, the results presented above for the S-L strategy implicitly include rewrites in the circuit-optimization stage. This led to fewer non-Pauli measurements but at the expense of increases in the maximum degree.

\section{Summary and Outlook}
\label{sec_outlook}

 QPatLib v1.0 was constructed to host MBQS patterns. The library has been initially populated with patterns executing Pauli string unitaries. For $n_q<6$ all possible Pauli string unitaries were converted into patterns. For $n_q\geq 6$, the Pauli strings were chosen from a set of benchmark molecular electronic-structure Hamiltonians with increasing numbers of active orbitals. Concatenating patterns effectively time evolves \cite{LEE2022} the Hamiltonians. Time evolution (unitary propagation) is an important component in hybrid circuit-based algorithms that can be adapted to MBQS. Future versions of the library will include new commuting-subset strategies, measurement patterns for Pauli string unitaries on other models (e.g., Hubbard models), and patterns for algorithms that do not use Pauli string unitaries.

Pauli strings were arranged into commuting subsets based on three example strategies: O-O, S-L, and Full. Changing strategies allows growth in pattern size while preserving the same unitary. The strategies discussed here reveal that pattern properties can be tuned effectively by adjusting strategies. For example, the S-L strategy increases the maximum graph degree over the O-O strategy. We also find that the Full strategy shows a rapid increase in the maximum layer depth with molecule complexity, whereas the maximum layer depth is bounded with the O-O strategy. More subset strategies are possible \cite{HAGBERG2008} and will be added to the library.

Lowering costs (node and edge count) is vital to implementing MBQS on near-term devices and constructing error-correction protocols. However, constraints may arise, e.g., from needs based on MBQC error-correction protocols or hardware-design limitations. Non-trivial constraints must therefore be imposed on parameters while compactifying patterns: graph degree, non-Pauli measurement counts, pattern width, pattern depth, and the maximum layer distance. Multivariable functionals can be designed to lower costs while imposing constraints in algorithms. Example large-pattern compactification algorithms include a flow-based routines \cite{PIUS2015}, simulated annealing \cite{KALDENBACH2025a,SHARMA2026} or machine learning \cite{LI2025}. The patterns defined here can be used to optimize constraint-aware compactification strategies.

\section{Data and Code Availability} 
\label{sec_data_availability}

The MBQS\textendash Pattern datasets used here are publicly available in QPatLib (version 1.0.0) at Zenodo under DOI 
\href{https://doi.org/10.5281/zenodo.20115266}{doi.org/10.5281/zenodo.20115266}. Quantum circuits are also stored here. Pattern schemas, benchmarking scripts, and analysis notebooks are available at GitHub, archived in the Zenodo QPatLib Community under the MIT License and DOI 
\href{https://doi.org/10.5281/zenodo.20114339}{doi.org/10.5281/zenodo.20114339}.  Additional intermediate data generated during the current study are available within the article figures or via the Zenodo records.

\begin{acknowledgments}
This material is based upon work supported by the U.S. Department of Energy, Office of Science, Office of Basic Energy Sciences Energy Frontier Research Centers program under Award Number DE-SC-0026289.
\end{acknowledgments}

\clearpage
\appendix

\section{Encoded Hamiltonian for \ce{Be2} }
\label{sec_encoded_Be2}

\begin{table}[h!]
\centering
\footnotesize
\begin{tabular}{|| @{\hspace{1pt}} c @{\hspace{3pt}} c @{\hspace{1pt}} c @{\hspace{-0pt}} || c @{\hspace{3pt}} c @{\hspace{1pt}} c @{\hspace{1pt}} ||}
\hline
$s$ & $c_s$ & String & $s$ & $c_s$ & String \\
\hline\hline
0  & -28.486511   & IIIIII & 31 & -0.011922474 & IXXYYI \\
1  & -0.015746339 & XXYYII & 32 & 0.011922474  & IXYYXI \\
2  & -0.010562348 & XXIIYY & 33 & 0.024660238  & IXZZZX \\
3  & 0.015746339  & XYYXII & 34 & 0.0039184745 & IXZZIX \\
4  & 0.010562348  & XYIIYX & 35 & 0.013184683  & IXZIZX \\
5  & 0.011922474  & XZXXZX & 36 & 0.0012622097 & IXIZZX \\
6  & 0.011922474  & XZXYZY & 37 & 0.011922474  & IYXXYI \\
7  & 0.024660238  & XZZZXI & 38 & -0.011922474 & IYYXXI \\
8  & 0.0039184745 & XZZZXZ & 39 & 0.024660238  & IYZZZY \\
9  & 0.0012622097 & XZZIXI & 40 & 0.0039184745 & IYZZIY \\
10 & 0.013184683  & XZIZXI & 41 & 0.013184683  & IYZIZY \\
11 & 0.014131832  & XIZZXI & 42 & 0.0012622097 & IYIZZY \\
12 & 0.015746339  & YXXYII & 43 & 0.2727863    & IZIIII \\
13 & 0.010562348  & YXIIXY & 44 & 0.064444304  & IZZIII \\
14 & -0.015746339 & YYXXII & 45 & 0.048697965  & IZIZII \\
15 & -0.010562348 & YYIIXX & 46 & 0.063396888  & IZIIZI \\
16 & 0.011922474  & YZYXZX & 47 & 0.052834541  & IZIIIZ \\
17 & 0.011922474  & YZYYZY & 48 & -0.01454893  & IIXXYY \\
18 & 0.024660238  & YZZZYI & 49 & 0.01454893   & IIXYYX \\
19 & 0.0039184745 & YZZZYZ & 50 & 0.01454893   & IIYXXY \\
20 & 0.0012622097 & YZZIYI & 51 & -0.01454893  & IIYYXX \\
21 & 0.013184683  & YZIZYI & 52 & 0.1951681    & IIZIII \\
22 & 0.014131832  & YIZZYI & 53 & 0.063734658  & IIZZII \\
23 & 0.2727863    & ZIIIII & 54 & 0.04490398   & IIZIZI \\
24 & 0.014131832  & ZXZZZX & 55 & 0.059452911  & IIZIIZ \\
25 & 0.014131832  & ZYZZZY & 56 & 0.1951681    & IIIZII \\
26 & 0.080913324  & ZZIIII & 57 & 0.059452911  & IIIZZI \\
27 & 0.048697965  & ZIZIII & 58 & 0.04490398   & IIIZIZ \\
28 & 0.064444304  & ZIIZII & 59 & 0.14970144   & IIIIZI \\
29 & 0.052834541  & ZIIIZI & 60 & 0.059608668  & IIIIZZ \\
30 & 0.063396888  & ZIIIIZ & 61 & 0.14970144   & IIIIIZ \\
\hline
\end{tabular}
\caption{Indexed list of coefficients and Pauli strings from the \ce{Be2} Hamiltonian encoded into 6 qubits from HamLib \cite{SAWAYA2024}. These entries determine the \ce{Be2} Hamiltonian using Eq.~\eqref{eq_Hamiltonian_string}. The Pauli strings act on qubits 0--5.}
\label{tab:be2-pauli-strings}
\end{table}

This section reproduces an example encoded Hamiltonian \cite{SAWAYA2024} to show how it is used in pattern construction. Table~\ref{tab:be2-pauli-strings} shows the components of the 6-qubit encoded Hamiltonian used in Eq.~\eqref{eq_Hamiltonian_string} for \ce{Be2}. The columns show the Hamiltonian coefficient and the corresponding Pauli string. Table~\ref{tab:be2-pauli-strings} is equivalent to an O-O strategy if each row defines a subset. Table~\ref{tab:be2-subsets} shows the string ordering of the S-L strategy using the strings from Table~\ref{tab:be2-pauli-strings}.

\begin{table}[ht!]
\centering
\footnotesize
\begin{tabular}{|| c l ||}
\hline
$\ell$ & Pauli strings in subset $\ell$ \\
\hline\hline
0 & \parbox[t]{0.85\columnwidth}{\raggedright
IIIIII, ZIIIII, ZZIIII, ZIZIII, ZIIZII, ZIIIZI, ZIIIIZ,
IZIIII, IZZIII, IZIZII, IZIIZI, IZIIIZ,
IIZIII, IIZZII, IIZIZI, IIZIIZ,
IIIZII, IIIZZI, IIIZIZ,
IIIIZI, IIIIZZ, IIIIIZ
} \\

1 & \parbox[t]{0.85\columnwidth}{\raggedright
ZXZZZX, ZYZZZY, IXZZZX, IXZZIX, IXZIZX, IXIZZX,
IYZZZY, IYZZIY, IYZIZY, IYIZZY
} \\

2 & \parbox[t]{0.85\columnwidth}{\raggedright
XXIIYY, XYIIYX, XZZZXI, XZZIXI, XZIZXI,
YXIIXY, YYIIXX, YZZZYI, YZZIYI, YZIZYI
} \\

3 & \parbox[t]{0.85\columnwidth}{\raggedright
XZXYZY, XZZZXZ, YXXYII, YIZZYI, IXXYYI, IIXYYX
} \\

4 & \parbox[t]{0.85\columnwidth}{\raggedright
XYYXII, XIZZXI, YZYXZX, YZZZYZ, IYYXXI, IIYXXY
} \\

5 & \parbox[t]{0.85\columnwidth}{\raggedright
XZXXZX, YZYYZY, IXYYXI, IYXXYI
} \\

6 & \parbox[t]{0.85\columnwidth}{\raggedright
XXYYII, YYXXII, IIXXYY, IIYYXX
} \\
\hline
\end{tabular}
\caption{The same Pauli strings as in Table~\ref{tab:be2-pauli-strings} but rearranged into commuting subsets indexed with $\ell$ using the S-L strategy.}
\label{tab:be2-subsets}
\end{table}

\section{Computational Methods}
\label{sec_computational_mehods}

We list the computational methods and tools used to produce the code and the library. The benchmark molecule models discussed in the main text are from the HamLib benchmarking library \cite{SAWAYA2024}. The chosen models use the Jordan--Wigner fermion-to-qubit mapping for electronic-structure models of: \ce{Be2} (6 qubits), \ce{H2} (8 qubits), \ce{BH} (10 qubits), \ce{OH} (12 qubits), \ce{NH} (14 qubits), \ce{C2} (18 qubits), \ce{N2} (22 qubits), and \ce{Na2} (24 qubits). The S-L subset strategy was constructed using the Networkx v2.8 package smallest-last \cite{HAGBERG2008}. Qiskit v2.2.3 \cite{JAVADI-ABHARI2024}, with optimization level 1 and default circuit layout, was used to transpile time-evolution unitaries of Pauli strings into quantum circuits using the following gate set: CNOT, S, H, $X$, $Y$, $Z$, SWAP, and RZ. Graphix v0.3.3 \cite{SUNAMI2022} was used to transpile quantum circuits into measurement patterns and prepare pattern schematics. The final code package is prepared as Python v3.11 Jupyter scripts. Sec.~\ref{sec_data_availability} lists locations for code and data availability.

Pattern properties are sensitive to these workflow choices. The high degree of variance in pattern properties implies the need for a standardized library. To see this, consider the S-L strategy applied to generate the $\ell=1$ subset of Pauli strings for \ce{Be2} (the associated Pauli strings are listed in Table~\ref{tab:be2-subsets}). Table~\ref{tab:qiskit_opt_ablation_long} compares circuit and pattern parameters that differ when the circuit-transpiler optimization level is changed. We see that changing the optimization level changes pattern properties even for small patterns. In defining the library, we choose Qiskit optimization level 1 \cite{JAVADI-ABHARI2024} to take advantage of simple rewrites while avoiding numerical compression of circuits performed with more aggressive optimization.

\begin{table}[t]
\vspace{0.2cm}
\centering
\begin{tabular}{||l|c|c||}
\hline
\multicolumn{3}{||c||}{Varying circuit optimization}\\[0.5ex]
\hline\hline
 & Opt.\ 0 & Opt.\ 1 \\
\hline
Circuit instructions & 162 & 115 \\
\hline
Node number $n$ & 290 & 266 \\
\hline
Max.\ degree $m_d$ & 4 & 6 \\
\hline
Pauli-$X$ measurements  & 234 & 210 \\
\hline
Pauli-$Y$ measurements  & 40 & 40 \\
\hline
Layers (causal flow) $n_l$ & 157 & 146 \\
\hline
Max edge layer span $m_{ld}$ & 154 & 143 \\
\hline
\end{tabular}
\caption{Results of varying transpiler optimization level (0 vs 1) \cite{JAVADI-ABHARI2024} for the simple example of the 6-qubit \ce{Be2} using the $\ell=1$ subset of Pauli strings in the S-L strategy (See Table~\ref{tab:be2-subsets}). The table shows that circuit instruction count and resulting measurement-pattern metrics (Table~\ref{tab_pattern_properties}) can vary with optimization even on a relatively small circuit.  QPatLib v1.0 chooses optimization level 1.}
\label{tab:qiskit_opt_ablation_long}
\end{table}

\section{Library Format}
\label{sec_library_format}

QPatLib contains two file types for each model: (i) the circuit files (QASM~3.0 format) define the circuit for the evolution operator for each subset; each file contains the circuit for one subset; and (ii) the pattern file (JSONL format) is a single file containing measurement patterns for all subsets. The file starts with a preamble containing the Hamiltonian definition in terms of Pauli strings \cite{SAWAYA2024} rearranged into subsets, provenance information, and results from the full pattern tests. Full pattern tests were only possible for $n_q\leq6$ due to computational limitations for larger patterns. The remainder of the file contains pattern entries, one entry per subset. Each entry starts with metadata for the subset and is followed by ASCII text for the pattern. The metadata include pattern properties as well as node-layer lists. The pattern is written in the standard defined by Eq.~\eqref{eq_pattern_convention} and includes signal shifting \cite{DANOS2007a}. The coefficients $c_s$ are symbolic in all patterns. The following shows an example subset entry for the pattern instance associated with subset $32$ for \ce{Be2} using the O-O strategy as well as a high level file schema for pattern files.

\begin{widetext}
\vspace{0.5cm}

\noindent
\begin{minipage}[t]{0.54\textwidth}
\footnotesize
\textbf{Example JSONL entry for the subset pattern in Fig.~\ref{fig_example_pattern_32_IXYYXI}:}

\vspace{2mm}
\begin{jsonverbatim}
{"Commuting subset": 32, "meta": {"node number": 44, "input_nodes": [0, 1, 2, 3, 4, 5], "output_nodes": [0, 28, 35, 42, 43, 5], "max degree": 4, "number Pauli X measurements": 29, "number Pauli Y measurements": 8, "number layers (causal flow)": 21, "max edge layer span": 14, "node_layer_list_causal_flow": {"0": 0, "1": 14, "2": 19, "3": 20, "4": 15, "5": 0, "6": 13, "7": 18, "8": 17, "9": 16, "10": 15, "11": 14, "12": 19, "13": 18, "14": 17, "15": 16, "16": 15, "17": 1, "18": 14, "19": 7, "20": 13, "21": 8, "22": 12, "23": 11, "24": 10, "25": 9, "26": 2, "27": 1, "28": 0, "29": 6, "30": 5, "31": 4, "32": 3, "33": 2, "34": 1, "35": 0, "36": 6, "37": 5, "38": 4, "39": 3, "40": 2, "41": 1, "42": 0, "43": 0}}, "pattern_ascii": "N(6) N(7) N(8) N(9) N(10) N(11) N(12) N(13) N(14) N(15) N(16) N(17) N(18) N(19) N(20) N(21) N(22) N(23) N(24) N(25) N(26) N(27) N(28) N(29) N(30) N(31) N(32) N(33) N(34) N(35) N(36) N(37) N(38) N(39) N(40) N(41) N(42) N(43) E(17,43) E(29,21) E(40,41) E(8,7) E(41,42) E(32,33) E(14,15) E(10,11) E(29,30) E(22,23) E(16,15) E(3,12) E(18,19) E(17,4) E(24,25) E(27,28) E(19,36) E(1,6) E(24,23) E(13,14) E(36,37) E(17,18) E(21,22) E(19,29) E(33,34) E(38,39) E(16,18) E(20,21) E(26,27) E(19,20) E(2,7) E(40,39) E(25,26) E(32,31) E(6,22) E(9,10) E(8,9) E(26,21) E(11,20) E(30,31) E(37,38) E(12,13) E(34,35) E(17,36) M(1) M(2,pi/2) M(7) [M(8)]{7} [M(9,pi/2)]{8,2} M(10) M(3,pi/2) M(12) [M(13)]{12} [M(14,pi/2)]{3,13} M(15) M(4) [M(16)]{3,13,15} M(18) [M(11)]{8,10,2} M(20) [M(6)]{1} M(22) [M(23,-2.0 * c[32])]{1,2,3,4,8,10,13,15,18,20,22} M(24) [M(25)]{1,2,3,4,8,10,13,15,18,20,22,24} M(26) [M(27)]{24,1,26,22} [M(21)]{2,3,4,8,10,13,15,18,20} M(29) [M(30,-pi/2)]{2,8,10,20,29} M(31) [M(32)]{2,8,10,20,29,31} [M(33,-pi/2)]{32,7,9,11,21,23,25,30} M(34) [M(19)]{3,4,13,15,18} M(36) [M(37,-pi/2)]{3,36,13,15,18} M(38) [M(39)]{3,36,38,13,15,18} [M(40,-pi/2)]{37,39,12,14,16,19,21,23,25} M(41) [M(17)]{4} Z(28,{24,1,26,22}) Z(35,{33,2,8,10,20,29,31}) Z(42,{3,36,38,40,13,15,18}) Z(43,{4}) X(28,{25,27,6,23}) X(35,{32,34,7,9,11,21,23,25,30}) X(42,{37,39,41,12,14,16,19,21,23,25}) X(43,{17,19,21,23,25})"}
\end{jsonverbatim}
\end{minipage}
\hfill
\begin{minipage}[t]{0.45\textwidth}
\footnotesize
\textbf{QPatLib pattern file schema (JSONL):}

\vspace{2mm}
\begin{description}
  \item[\textbf{Format}] One JSON object per line. The file begins with a header (Hamiltonian-level metadata), may include optional summary/comparison blocks, followed by one entry per commuting subset.
  \item[\textbf{Header}] Hamiltonian name; instance tag; number of qubits; full list of Pauli-string coefficients and corresponding Pauli terms; subset/coloring strategy; mapping from subset index to commuting Pauli strings; provenance (software versions, backend, flags, seed); global summary statistics (e.g., concatenated depth, global max degree).
  \item[\textbf{Summary}] Strategy-comparison block(s) that report aggregate pattern statistics for an alternative construction (e.g., \texttt{full\_hamiltonian}): total node count, max degree, depth, max edge layer span, and Pauli-measurement counts.
  \item[\textbf{Subset}] For each ``Commuting subset'':\par
  \textit{meta}: node number, input/output node lists, max degree, Pauli-$X$ and Pauli-$Y$ measurement counts, number of layers (flow depth), max edge layer span, and a node$\rightarrow$layer map (e.g., causal-flow layering).\par
  \textit{pattern\_ascii}: the measurement-calculus string (nodes, edges, measurements with signal dependencies, and byproduct operators) stored as readable ASCII using variables/symbols, e.g., $\pi$ and $c[s]$.
\end{description}
\end{minipage}

\end{widetext}

\section{Validation}
\label{sec_validation}

The patterns were validated using several methods. Code accompanying the library can be used to reproduce the automated tests. First, all measurement patterns were compared against the corresponding quantum circuit. Both the circuit and the pattern were supplied with a random input wavefunction. The outputs were compared using the Bhattacharyya coefficient. All patterns stored in the library passed with a coefficient $> 0.999$. Second, the concatenation procedure for the O-O and S-L strategies was checked against the Full strategy for $n_q\leq 6$. The Full strategy could not be checked for patterns for molecules with a larger number of qubits due to the pattern size. To check patterns with $n_q\leq 6$, a tensornetwork simulator \cite{SUNAMI2022} was used to compute the wavefunctions obtained by concatenating subsets. 
As a quick spot-check we compared the first 5 wavefunction amplitudes of concatenated O-O and S-L patterns were checked against the first 5 amplitudes of the wavefunction for the Full strategy to within 6 significant digits. Then the overlap of the wavefunctions was checked to be unity to within 6 significant digits. Third, all patterns stored in the database were checked to satisfy causal flow using Graphix \cite{SUNAMI2022}.

\newpage
\bibliographystyle{quantum}
\bibliography{references}

\end{document}